\pdfoutput=1
\documentclass[11pt]{article}
\usepackage{acl}
\usepackage{times}
\usepackage{latexsym}
\usepackage[T1]{fontenc}
\usepackage[utf8]{inputenc}
\usepackage{microtype}
\usepackage{subcaption}
\usepackage{booktabs}
\usepackage{amsmath, amsthm, amssymb, amsfonts,cleveref}
\usepackage{graphicx,tikz}

\usepackage{todonotes}

\usetikzlibrary{arrows,decorations.markings,decorations.pathreplacing,calligraphy,spy}
\def\sectionautorefname~#1\null{\S#1\null}
\def\subsectionautorefname~#1\null{\S#1\null}
\def\equationautorefname~#1\null{(#1)\null}
\title{A Deep Metric Learning Approach to Account Linking} 
\author{Aleem Khan, Elizabeth Fleming,
Noah Schofield, Marcus Bishop, \and Nicholas Andrews \\
Human Language Technology Center of Excellence\\
Johns Hopkins University\\
{\tt \{akhan141,eflemin9,nschofi1,marcus.bishop,noa\}@jhu.edu}
}

\begin{document}
\maketitle
\begin{abstract}
We consider the task of linking social media accounts 
that belong to the same author in an automated fashion on the basis of the content and metadata of their corresponding document streams.
We focus on learning an embedding that maps variable-sized samples of user activity---ranging from single posts to entire months of activity---to a vector space, where 
samples by the same author map to nearby points.
The approach does not require human-annotated data for training purposes, which 
allows us to leverage large amounts of social media 
content. 
The proposed model outperforms several competitive baselines
under a novel evaluation framework modeled after established recognition benchmarks in other domains.
Our method achieves high linking accuracy, 
even with small samples from accounts not seen at training time, a prerequisite for practical applications of the proposed linking framework.
\end{abstract}


\section{Introduction}

\begin{figure*}[t!]
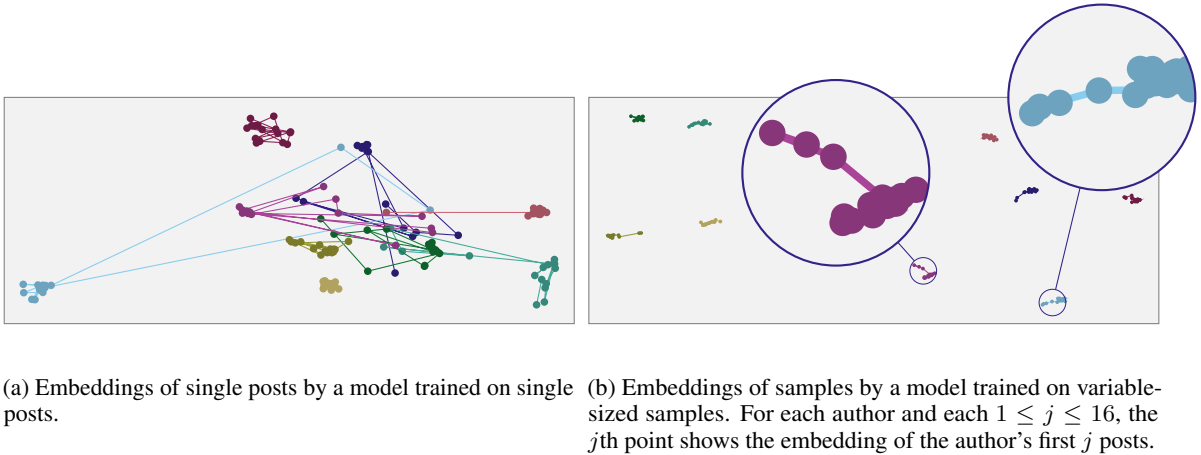

  \centering
  \begin{subfigure}[t]{7.5cm}
    \centering
    \input{single}
    \caption{Embeddings of single posts by a model trained on single posts.}
    \label{fig:single}
  \end{subfigure}%
  ~
  \begin{subfigure}[t]{7.5cm}
    \centering
    \input{attention}
    \caption{Embeddings of samples by a model trained on variable-sized samples. For each author and each $1\le j\le 16$, the $j$th point shows the embedding of the author's first $j$~posts.}
    \label{fig:attention}
  \end{subfigure}
  \caption{UMAP projections~\cite{mcinnes2020umap} of embeddings of document streams. Each image shows the same 16~Reddit posts by 9~randomly chosen authors, connected chronologically. Posts by each author are shown in the same color in both figures.}
\end{figure*}

The scale and anonymity of social media pose systematic challenges for manual moderation efforts~\cite{pennycook2020fighting,broniatowski2018weaponized}. These challenges have motivated the development of automated methods to identify abusive content, such as \newcite{davidson2017automated}, which considers automatically classifying hate speech, or \newcite{violentExtermeism}, which deals with detecting violent extremism, both in the Twitter domain. 

However, automatic moderation remains a difficult problem. Indeed, existing methods based on hand-constructed resources such as keyword lists may fail to adapt to novel trends~\cite{corbett2018measure} whereas automatic methods based on statistics of large corpora may exhibit harmful biases~\cite{caliskan2017semantics}. 
Additionally, individual posts may fail to contain sufficient information to reliably identify them as harmful.

This work considers \emph{account-level} moderation.
Specifically, we consider the problem of determining whether two {\em document streams} share the same author, based on {\em samples} from those streams rather than individual documents.
This capability has numerous applications, such as detecting users attempting to circumvent account bans, identifying sockpuppet accounts, and detecting coordinated disinformation campaigns involving multiple authors controlling multiple accounts.

As a motivating application, we consider the enforcement of account bans on anonymous platforms, such as Reddit. Given a new account, the problem is to automatically identify whether it matches any previously banned account, which amounts to making binary decisions about whether pairs of accounts share the same author. Variations of this problem have been studied before. For example, \newcite{schwartz2013authorship} learn a classifier to determine whether the author of a Twitter comment belongs to a small, closed set of authors. In contrast, we are interested in an open-world setting, requiring binary decisions about arbitrary pairs of accounts.
This introduces a number of challenges.

First, any individual comment may be too short to serve as the basis for linking accounts. \autoref{fig:single} illustrates this empirically using a variation of our model, where embeddings of individual comments from the same account fail to coalesce, making it difficult to assert that an account has the same author as another account. See~\autoref{sec:experiments} for further experimental details. Therefore, we focus on \emph{aggregating} information across contiguous sequences of documents. \autoref{fig:attention} illustrates the impact of aggregation using our full model, where aggregations of contiguous sequences of documents from the same account exhibit an approximate convergence behavior as the number of documents aggregated increases. 
In fact, the motivating application above {\em requires} linking accounts on the basis of samples of widely varying sizes. Indeed, banned accounts typically have many documents, all of which we would like to consider,
while new accounts generally have few documents, which nevertheless must be linked to banned accounts as quickly as possible to mitigate abusive behavior.

The second challenge is that of spurious associations. For example, over a short period of time, an author may discuss only a single, narrow topic. While a na\"{i}ve model based on word statistics alone might be sufficient to link such an account to another, this approach would fail to generalize over longer periods of time due to topic drift. 
Both our training procedure and evaluation framework have been designed to ensure that our model learns the appropriate invariances to identify same-authorship, rather than being correct for the wrong reasons~\cite{mccoy2019right}.
Namely, samples from an account are drawn from different time periods in each training iteration (see~\autoref{sec:sampling}), while the evaluation data consists of posts by accounts not seen at training time and is future to all the training data (see~\autoref{sec:ranking}).

Finally, the numbers of banned and new accounts may be quite large, requiring a still larger number of pairwise comparisons. For this reason, our proposed approach to account linking consists of embedding variable-sized samples from document streams into a metric space whereby samples likely to have been composed by the same author map to nearby points.
Under this embedding, comparisons between document streams amount to pairwise distance calculations, so our approach is highly scalable and amenable to various optimizations, such as approximate nearest neighbor methods.

Our primary contributions are the following:
\begin{itemize}
\item We provide a simple but effective data augmentation strategy which enables embedding variable-sized samples. In addition, we successfully train such an embedding on a large-scale dataset consisting of more than 300~million comments from 1~million distinct accounts using scalable losses.
\item We propose a novel framework to assess account linking performance
focused on challenging conditions and minimizing the impact of incidental authorship features, such as topic.   
In particular, we propose benchmark datasets as well as verification metrics tailored to our application.
\end{itemize}

Our code, data splits,
and scripts to reproduce our experiments are available at \url{http://github.com/noa/naacl2021}.
\section{Learning embeddings of document streams}\label{sec:model}

\begin{figure*}[ht!]
\begin{center}
\begin{tikzpicture}
\definecolor{localColor0}{RGB}{51,34,136}
\definecolor{localColor2}{RGB}{68,170,153}
\definecolor{localColor3}{RGB}{17,119,51}
\definecolor{localColor7}{RGB}{136,34,85}

\def\h{5.5}
\draw (0,\h) node[anchor=east] { $\left(t_1,r_1\right)$};
\draw [->,>=stealth] (0,\h) -- (1,\h) node[midway,above]{ $E$};
\draw[fill=localColor7!50] (1,\h-.2) rectangle (2,\h+.2);
\draw (1.5,\h)  node { $t_1$};
\draw[fill=localColor3!50] (2,\h-.2) rectangle (3,\h+.2);
\draw (2.5,\h)  node { $r_1$};
\draw[fill=localColor0!50] (3,\h-.2) rectangle node[anchor=north] (y1) {} (4,\h+.2);
\draw (3.5,\h)  node{ $x_1$};
\draw[fill=localColor2!50] (5,\h-.2) rectangle (8,\h+.2);

\def\h{5}
\draw (0,\h) node[anchor=east] { $\left(t_2,r_2\right)$};
\draw [->,>=stealth] (0,\h) -- (1,\h) node[midway,above]{ $E$};
\draw[fill=localColor7!50] (1,\h-.2) rectangle (2,\h+.2);
\draw (1.5,\h)  node { $t_2$};
\draw[fill=localColor3!50] (2,\h-.2) rectangle (3,\h+.2);
\draw (2.5,\h)  node { $r_2$};
\draw[fill=localColor0!50] (3,\h-.2) rectangle node[anchor=north] (y2) {} (4,\h+.2);
\draw (3.5,\h)  node{ $x_2$};
\draw[fill=localColor2!50] (5,\h-.2) rectangle (8,\h+.2);

\draw (-0.75,4.6) node[anchor=center] {$\vdots$};

\def\h{4}
\draw (0,\h) node[anchor=east] { $\left(t_M,r_M\right)$};
\draw [->,>=stealth] (0,\h) -- (1,\h) node[midway,above]{ $E$};
\draw[fill=localColor7!50] (1,\h-.2) rectangle (2,\h+.2);
\draw (1.5,\h)  node { $t_M$};
\draw[fill=localColor3!50] (2,\h-.2) rectangle (3,\h+.2);
\draw (2.5,\h)  node { $r_M$};
\draw[fill=localColor0!50] (3,\h-.2) rectangle node[anchor=north west] (yM) {} (4,\h+.2);
\draw (3.5,\h)  node{ $x_M$};
\draw[fill=localColor2!50] (5,\h-.2) rectangle node (c) {} (8,\h+.2);

\draw [decoration={calligraphic brace,amplitude=5pt},decorate, line width=2pt] (4,5.7) -- (4,3.8) node[midway] (a) {};
\draw [decoration={mirror,calligraphic brace,amplitude=5pt},decorate, line width=2pt] (5,5.7) -- (5,3.8) node[midway] (b) {};
\draw[->,>=stealth] (a) -- node[midway,above] {$A$} (b);

\draw [decoration={mirror,calligraphic brace,amplitude=5pt},decorate, line width=2pt] (8,4-.2) -- (8,5.5+.2) node[midway] (c) {};
\draw[fill=localColor2!50] (9,4.75-.2) rectangle node[anchor=north] (d) {} (12,4.75+.2);
\draw[->,>=stealth] (c) -- node[midway,above] {$M$} (9,4.75);
\draw[fill=localColor2!50] (9,3.75-.2) rectangle node[anchor=center] (e) {$f_\theta\left(a\right)$} (12,3.75+.2);
\draw[->,>=stealth] (d) -- node[midway,left] {$P$} (e);

\def\h{3}
\def\v{0}
\draw (\v,\h) node[anchor=east] (x) { $x_1$};
\node [draw,fill=localColor0!50,shape=rectangle,minimum height=20,minimum width=15,anchor=center] (b1) at (\v+1,\h) {};
\draw [->,>=stealth] (x) -- (b1) node[midway,above]{ $E$};
\node [draw,fill=localColor0!50,shape=rectangle,minimum height=20,minimum width=15,anchor=center] (b2) at (\v+2,\h) {};
\draw [->,>=stealth] (b1) -- (b2) node[midway,above]{ $C$};
\node [draw,fill=localColor0!50,shape=rectangle,minimum height=20,minimum width=5,anchor=center] (b3) at (\v+3,\h) {};
\draw [->,>=stealth] (b2) -- (b3) node[midway,above]{ $P$};
\draw [->,>=stealth] (b3) -- (y1);

\def\h{2}
\def\v{.75}
\draw (\v,\h) node[anchor=east] (x) { $x_2$};
\node [draw,fill=localColor0!50,shape=rectangle,minimum height=20,minimum width=15,anchor=center] (b1) at (\v+1,\h) {};
\draw [->,>=stealth] (x) -- (b1) node[midway,above]{ $E$};
\node [draw,fill=localColor0!50,shape=rectangle,minimum height=20,minimum width=15,anchor=center] (b2) at (\v+2,\h) {};
\draw [->,>=stealth] (b1) -- (b2) node[midway,above]{ $C$};
\node [draw,fill=localColor0!50,shape=rectangle,minimum height=20,minimum width=5,anchor=center] (b3) at (\v+3,\h) {};
\draw [->,>=stealth] (b2) -- (b3) node[midway,above]{ $P$};
\draw [->,>=stealth] (b3) -- (y2);

\draw (1.25,1.3125) node {$\ddots$};

\def\h{.5}
\def\v{2.5}
\draw (\v,\h) node[anchor=east] (x) { $x_M$};
\node [draw,fill=localColor0!50,shape=rectangle,minimum height=20,minimum width=15,anchor=center] (b1) at (\v+1,\h) {};
\draw [->,>=stealth] (x) -- (b1) node[midway,above]{ $E$};
\node [draw,fill=localColor0!50,shape=rectangle,minimum height=20,minimum width=15,anchor=center] (b2) at (\v+2,\h) {};
\draw [->,>=stealth] (b1) -- (b2) node[midway,above]{ $C$};
\node [draw,fill=localColor0!50,shape=rectangle,minimum height=20,minimum width=5,anchor=center] (b3) at (\v+3,\h) {};
\draw [->,>=stealth] (b2) -- (b3) node[midway,above]{ $P$};
\draw [->,>=stealth] (b3) -- (yM);

\end{tikzpicture}
\end{center}
\caption{Illustration of the proposed model architecture. Each action~$a_i$ consists of text content~$x_i\in\mathbb{Z}^\ell$, a subreddit feature~$r_i\in\mathbb{Z}$, and a publication time~$t_i\in\left\{0,1,\ldots,23\right\}$. These elements are combined as shown using various embedding lookups~$E$, one-dimensional convolutions~$C$, and linear projections~$P$, along with an attention mechanism~$A$ and a max pooling layer~$M$.}
\label{fig:arch}
\end{figure*}
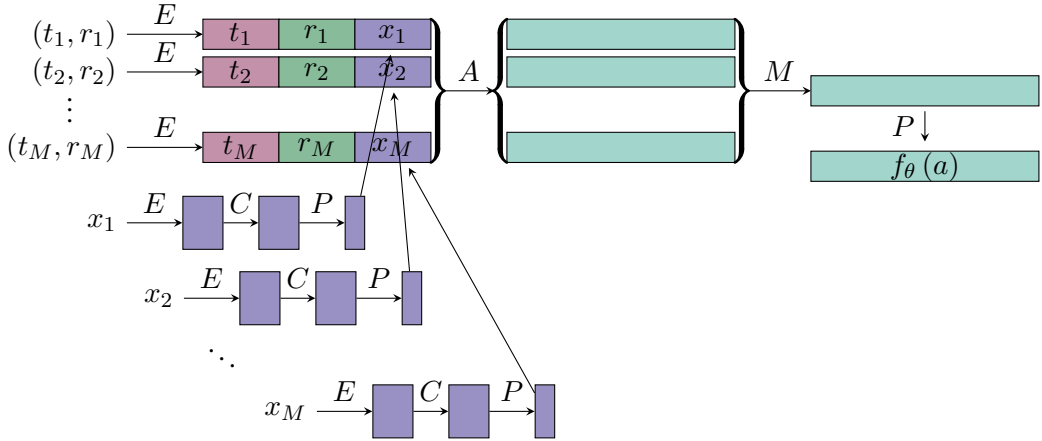

We treat a document stream as a sequence of timestamped actions $a_1,a_2,\ldots,a_L$ where each $a_i$ is a structure containing the data comprising an action. The possible contents of $a_i$ are specific to the document stream, but include at least a timestamp $t_i$ such that $t_1<t_2<\cdots<t_L$.

In this work we focus on textual content published on social media platforms, although the approach would easily extend to allow $a_1,a_2,\ldots,a_L$ to contain other modalities such as images or video, which would be handled similarly. In addition, we also avail of certain categorical features contained in $a_1,a_2,\ldots,a_L$ such as hashtags or the subreddit to which a comment was posted.

A {\em sample} from a document stream is a contiguous subsequence of its actions. We introduce an embedding $f_\theta$ in~\autoref{sec:arch} mapping a {\em variable-sized} sample to a point in a vector space, such that the Euclidean distance between the embeddings of two samples quantifies the likelihood that they belong to the same author.

\subsection{Architecture}\label{sec:arch}
We define an embedding $f_\theta$ as follows. This embedding is illustrated in~\autoref{fig:arch}.
Consider a sample $a=\left(a_1,a_2,\ldots,a_M\right)$ where each action~$a_i$ consists of a subreddit feature~$r_i$, a timestamp~$t_i$, and text content~$x_i$.\footnote{In addition to $r_i,t_i,x_i$ platforms like Reddit contain further metadata, such as the thread title and the submission and parent comments, all of which might help distinguish users because users read them and chose to respond. Incorporating these features would be interesting to explore in future work.}
We encode $t_i$ as the corresponding hour of the day and $r_i$ by lookup in the list of 2048~most common subreddits, resulting in $t_i\in\left\{0,1,\ldots,23\right\}$ and $r_i\in\left\{0,1,\ldots,2048\right\}$, where $r_i=2048$ when the subreddit is not among the top~2048.
We encode the text feature using the SentencePiece unigram subword model~\cite{kudo2018subword}, resulting in $x_i\in\mathbb{Z}^\ell$, where the parameter~$\ell$ is defined in~\autoref{sec:sample}. Note that the chosen vocabulary size impacts the amount of content that can be encoded with $\ell$~integers. Further details on the choice of text encoding are provided in~\autoref{sec:encode}.

We replace each token of $x_i$ with a corresponding learned embedding in $\mathbb{R}^N$ for all $1\le i\le M$, resulting in $M$~matrices in $\mathbb{R}^{\ell\times N}$.
We apply one-dimensional convolutions of widths 2, 3, and 4 along the first axis of each, concatenate the convolved matrices along their second axes, max-pool along the first axis, and concatenate the results with learned embeddings of the corresponding subreddit features and one-hot encodings of the corresponding time features, resulting in $M$~vectors, which we aggregate using dot-product attention. The resulting sequence of vectors is projected to a single vector through max-pooling followed by two fully-connected layers with bias, resulting in the encoding $f_\theta\left(a\right)\in\mathbb{R}^D$ of the sample~$a$.

\subsection{Text Sampling}\label{sec:sample}
 The variance in the lengths of documents poses computational challenges when aggregating large samples. Therefore we resort to truncating each document to a fixed number~$\ell$ of tokens, padding any documents containing fewer than $\ell$~tokens.
We take $\ell=32$ after observing that Reddit posts have an average length of approximately 43~tokens (see~\autoref{sec:mud}). 
 
We also experimented with a more complicated text sampling strategy, namely
sampling contiguous segments of $\ell$~tokens from each post uniformly at random during training.
While this approach leverages all available textual information by affording slightly different samples of each post in each iteration of training, we found it to  yield similar results and also complicates the comparison to our primary baseline model, which uses the prefix method described above.


\subsection{Sample selection}\label{sec:sampling}
\begin{figure}
    \centering
    \begin{tikzpicture}[domain=0:1,samples=100,xscale=4]
\definecolor{localColor0}{RGB}{51,34,136}
\definecolor{localColor3}{RGB}{17,119,51}
\definecolor{localColor7}{RGB}{136,34,85}
\draw[ystep=0.5,xstep=0.125,very thin,color=gray] (-0.025,-0.1) grid (1.025, 3.1); 
\draw[thick,->] (-0.05,0) -- (1.05,0) node[right] {$x$};
\draw[thick,->] (0,-0.2) -- (0,3.2) node[above] {$y$};
\draw[color=localColor0,thick] plot (\x,1) node[right] {$\mathrm{Unif}\left(0,1\right)$};
\draw[color=localColor7,thick] plot (\x, {3*\x*\x}  ) node[right] {$\mathrm{Beta}\left(3,1\right)$};
\foreach \y in {1,2,3}
  \draw (0,\y) node[left] {$\y$};
\draw (0.5,0) node[below] {$\frac{1}{2}$};
\draw (1,0) node[below] {$1$};
\end{tikzpicture}
    \caption{In contrast with choosing sample sizes uniformly at random, sampling according to a left skewed Beta distribution results in more samples of sizes closer to the specified maximum.}
    \label{fig:beta}
\end{figure}
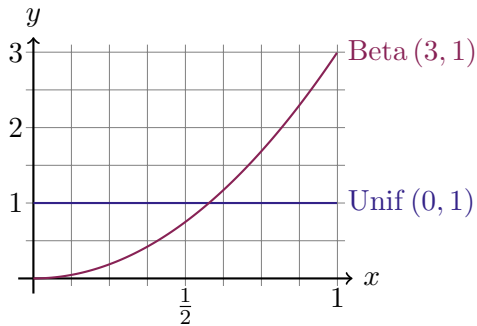
During training we randomly select document stream samples of sizes varying between $R=1$ and $S=16$, which we regard as hyperparameters of the model. To select a sample from the stream $a_1,a_2,\ldots,a_L$ we first choose its length $M=R+\left\lceil x\left(S-R\right)\right\rceil$ where $x\sim\mathrm{Beta}\left(3,1 \right)$ and take the sample
$a_i,a_{i+1},\ldots,a_{i+M-1}$ 
where $1\le i\le L-M+1$ is chosen uniformly at random.
Selecting $M$ according to $\mathrm{Beta}\left(3,1\right)$ provides an expected sample size closer to~$S$ than to~$R$, a trade-off that allows the model to quickly learn features of a document stream by exposing it to larger samples most of the time, while still maintaining the flexibility to handle samples of varying sizes. Indeed, the latter is critical in the evaluation described in~\autoref{sec:al}, which requires linking large samples to small samples.
The density function of $\mathrm{Beta}\left(3,1\right)$ is shown in~\autoref{fig:beta} together with that of the uniform distribution $\mathrm{Unif}\left(0,1\right)$ for comparison. We explore the benefits of $\mathrm{Beta}\left(3,1\right)$ and other related distributions in~\autoref{sec:beta_expt}. 

\subsection{Scalable deep metric learning losses}\label{sec:loss} 

Deep metric learning methods aim to embed observations into a low-dimensional space such that instances from the same class map to nearby points under a chosen metric, such as Euclidean distance. In our setting, we take the instances to be document stream samples and the classes to be the corresponding accounts, which serve as proxies for latent authorship. Therefore, training the mapping $f_\theta$ defined in~\autoref{sec:arch} using metric learning affords an embedding under which samples by the same author map to nearby points.

Recent work in deep metric learning has introduced a number of training objectives with state of the art performance on computer vision tasks~\cite{kim2020proxy,wang2019crossbatch}. Unfortunately, many of these objectives scale linearly with the number~$K$ of classes considered due to a costly linear projection onto $\mathbb{R}^K$.
Note that because account names in effect provide labels for the corresponding document streams, we may use raw social media content to fit our model directly, availing of a virtually unlimited source of data.
We stipulate that the ability to exploit larger amounts of data may be more important than per-example efficiency, and therefore consider the classical {\em triplet loss} \cite{schroff2015facenet} in our experiments, whose complexity does not depend on~$K$. In particular, we use semi-hard negative mining with a fixed margin penalty.
We also consider the {\em top-$k$} loss recently proposed
by~\citet{Lu_2019_ICCV}, which optimizes precision-at-$k$ as follows. Given targets ranked by similarity to a query, top-$k$ arranges for as many matches as possible to be among the top~$k$ ranked targets. It accomplishes this by penalizing only those targets that would need to move the smallest amount in order to maximize the number of matching targets among the top~$k$. Like triplet loss, top-$k$ also uses an additive margin penalty to separate classes.
See~\autoref{sec:hyper} for further experimental details on both loss functions.

\section{Related work}\label{sec:relatedWork}

The separate but related problem of \emph{closed-world} author attribution has received considerable attention. For example, the PAN 2019 challenge~\cite{pan:2019b} employed a closed-world setting with a small number of authors that are the same at training and test time. That task also considered longer documents, obviating the need for aggregating evidence of authorship across multiple documents.

Generic text embedding methods such as the universal sentence encoder~\cite{cer2018universal} and BERT~\cite{devlin2018bert} are fit using auxiliary tasks, such as conditional language modeling. In the case of BERT, this is usually followed by supervised fine-tuning for a downstream task of interest. In this work, we are interested in learning representations that are immediately useful for our account linking task. 
However, because a large corpus of task-specific training data may be collected without human supervision, the benefits of generative pre-training are diminished in our setting. Indeed, the parameters of the text encoding are learned from a random initialization in all our experiments.
Our approach is further distinguished from generic embedding methods by featuring a \emph{multi-document} embedding, mapping a sequence of documents to a single vector, where each document may consist of both text and metadata.

The most closely related prior work is the Invariant User Representation (IUR) proposed by~\citet{andrews-bishop-2019-learning}, whose approach is broadly similar to ours, but only considers samples of a fixed size. Our approach may be viewed as a generalization of that work in support of the account linking task. In addition we use a simpler dot product attention mechanism and introduce the use of scalable metric learning losses in~\autoref{sec:loss}, which enable us to train our model on an order of magnitude more data than previously considered. We validate these improvements in~\autoref{sec:ranking} using the ranking task proposed by~\citet{andrews-bishop-2019-learning}. We also adapt IUR to serve as a baseline in our primary linking task in~\autoref{sec:al}.

We believe that our treatment of account linking as a pairwise recognition task between document stream samples and our proposed general-purpose evaluation protocols are both novel.
However, in prior work, there have been several platform-specific approaches to account linking. For example~\citet{silvestri2015linking} explore a heuristic approach to linking accounts across social media platforms.
Separately, on platforms with rich social network information, graph-matching methods have been explored~\cite{fan2012graph}. Our focus is on content-based account linking, which is more general than  prior methods we are aware of. Some other related but distinct problems include detecting deceptive accounts~\cite{van2018using} and authorship classification of short messages~\cite{ishihara2011forensic}.
\section{Experiments}\label{sec:experiments}

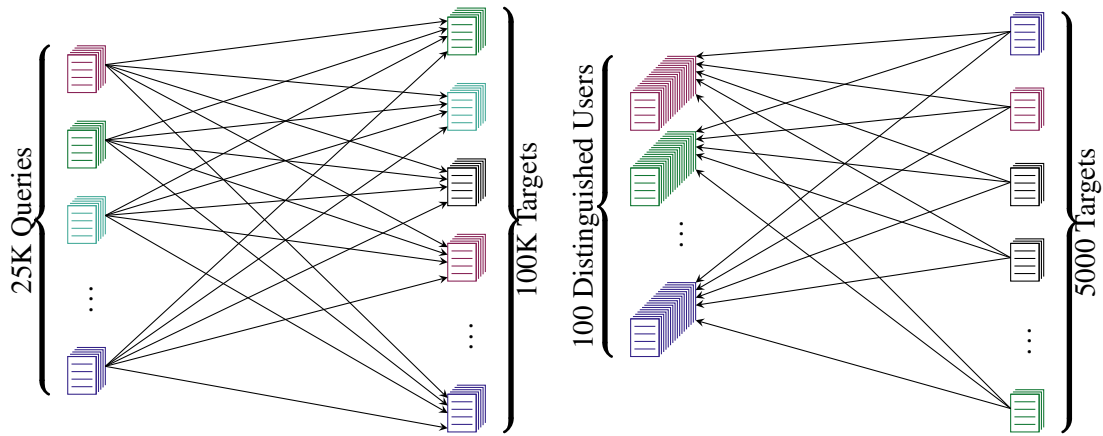
\begin{figure*}[t!]
  \centering
  \begin{subfigure}[t]{0.45\textwidth}
    \centering
    \definecolor{localColor0}{RGB}{51,34,136}
\definecolor{localColor1}{RGB}{136,204,238}
\definecolor{localColor2}{RGB}{68,170,153}
\definecolor{localColor3}{RGB}{17,119,51}
\definecolor{localColor4}{RGB}{153,153,51}
\definecolor{localColor5}{RGB}{221,204,119}
\definecolor{localColor6}{RGB}{204,102,119}
\definecolor{localColor7}{RGB}{136,34,85}
\definecolor{localColor8}{RGB}{170,68,153}

\def\nPages{4}
\def\pageThickness{0.03}
\tikzset{
  stack/.pic = {
    \draw (0,0) rectangle (0.375,0.5);
    \foreach \x in {0.1,0.2,0.3,0.4}
      \draw (0.05,\x) -- (0.325,\x);
    \foreach \x in {0,...,\nPages} {
      \def\d{\pageThickness*\x};
      \draw (\d,0.5+\d) -- (.375+\d,0.5+\d) -- (.375+\d,\d);
    }
  }
}

\begin{tikzpicture}
\def\stackHeight{\nPages*\pageThickness+0.5}
\def\midpointH{\stackHeight/2}
\def\stackWidth{\nPages*\pageThickness+0.375}
\def\midpointW{\stackWidth/2}

\begin{scope}[color=localColor0] \pic at (0,0.5) {stack}; \end{scope}
\begin{scope}[color=localColor2] \pic at (0,2.5) {stack}; \end{scope}
\begin{scope}[color=localColor3] \pic at (0,3.5) {stack}; \end{scope}
\begin{scope}[color=localColor7] \pic at (0,4.5) {stack}; \end{scope}
\node at (\midpointW,1.5+\midpointH) {$\vdots$};
\begin{scope}[color=localColor0] \pic at (5,0) {stack}; \end{scope}
\begin{scope}[color=localColor7] \pic at (5,2) {stack}; \end{scope}
\begin{scope}[color=black] \pic at (5,3) {stack}; \end{scope}
\begin{scope}[color=localColor2] \pic at (5,4) {stack}; \end{scope}
\begin{scope}[color=localColor3] \pic at (5,5) {stack}; \end{scope}
\node at (\midpointW+5,1+\midpointH) {$\vdots$};
\foreach \x in {0.5,2.5,3.5,4.5}{
  \foreach \y in {0,2,3,4,5}{
    \draw[->,>=stealth] (\stackWidth,\x+\midpointH) -- (5,\y+\x/10);
  }
}
\draw [decoration={calligraphic brace,amplitude=5pt},decorate, line width=2pt] (-0.25,0.5) -- (-0.25,4.5+\stackHeight) node[midway,rotate=90,above] {25K Queries};
\draw [decoration={calligraphic brace,mirror,amplitude=5pt},decorate,line width=2pt] (5.25+\stackWidth,0) -- (5.25+\stackWidth,5+\stackHeight) node[midway,rotate=90,below] {100K Targets};
\end{tikzpicture}
    \caption{Ranking evaluation: for each query, the targets are ranked by similarity to the query.}
    \label{fig:ranking}
  \end{subfigure}%
  ~
  \begin{subfigure}[t]{0.45\textwidth}
    \centering
    \definecolor{localColor0}{RGB}{51,34,136}
\definecolor{localColor1}{RGB}{136,204,238}
\definecolor{localColor2}{RGB}{68,170,153}
\definecolor{localColor3}{RGB}{17,119,51}
\definecolor{localColor4}{RGB}{153,153,51}
\definecolor{localColor5}{RGB}{221,204,119}
\definecolor{localColor6}{RGB}{204,102,119}
\definecolor{localColor7}{RGB}{136,34,85}
\definecolor{localColor8}{RGB}{170,68,153}
\def\nPagesL{16}
\def\pageThickness{0.03}
\tikzset{
  stackBig/.pic = {
    \draw (0,0) rectangle (0.375,0.5);
    \foreach \x in {0.1,0.2,0.3,0.4}
      \draw (0.05,\x) -- (0.325,\x);
    \foreach \x in {0,...,\nPagesL} {
      \def\d{\pageThickness*\x};
      \draw (\d,0.5+\d) -- (.375+\d,0.5+\d) -- (.375+\d,\d);
    }
  }
}

\def\nPagesR{2}
\tikzset{
  stackSmall/.pic = {
    \draw (0,0) rectangle (0.375,0.5);
    \foreach \x in {0.1,0.2,0.3,0.4}
      \draw (0.05,\x) -- (0.325,\x);
    \foreach \x in {0,...,\nPagesR} {
      \def\d{\pageThickness*\x};
      \draw (\d,0.5+\d) -- (.375+\d,0.5+\d) -- (.375+\d,\d);
    }
  }
}

\begin{tikzpicture}
\def\stackHeightL{\nPagesL*\pageThickness+0.5}
\def\midpointHL{\stackHeightL/2}
\def\stackWidthL{\nPagesL*\pageThickness+0.375}
\def\midpointWL{\stackWidthL/2}
\def\stackHeightR{\nPagesR*\pageThickness+0.5}
\def\midpointHR{\stackHeightR/2}
\def\stackWidthR{\nPagesR*\pageThickness+0.375}
\def\midpointWR{\stackWidthR/2}

\begin{scope}[color=localColor0] \pic at (0,1) {stackBig}; \end{scope}
\begin{scope}[color=localColor3] \pic at (0,3) {stackBig}; \end{scope}
\begin{scope}[color=localColor7] \pic at (0,4) {stackBig}; \end{scope}
\node at (\midpointWL,2+\midpointHL) {$\vdots$};
\begin{scope}[color=localColor3] \pic at (5,0) {stackSmall}; \end{scope}
\begin{scope}[color=black] \pic at (5,2) {stackSmall}; \end{scope}
\begin{scope}[color=black] \pic at (5,3) {stackSmall}; \end{scope}
\begin{scope}[color=localColor7] \pic at (5,4) {stackSmall}; \end{scope}
\begin{scope}[color=localColor0] \pic at (5,5) {stackSmall}; \end{scope}
\node at (\midpointWR+5,1+\midpointHR) {$\vdots$};
\foreach \x in {1,3,4}{
  \foreach \y in {0,2,3,4,5}{
    \draw[->,>=stealth] (5,\y+\midpointHR) -- (\stackWidthL,\x+\y/10+\nPagesL*\pageThickness);
  }
}
\draw [decoration={calligraphic brace,amplitude=5pt},decorate, line width=2pt] (-0.25,1) -- (-0.25,4+\stackHeightL)
node[midway,above,rotate=90] {100 Distinguished Users};
\draw [decoration={calligraphic brace,mirror,amplitude=5pt},decorate,line width=2pt] (5.25+\stackWidthR,0) -- (5.25+\stackWidthR,5+\stackHeightR) node[midway,below,rotate=90] {5000 Targets};
\end{tikzpicture}
    \caption{Linking evaluation: for each target, the queries are determined to match each query or not.}
    \label{fig:linking}
  \end{subfigure}
  \caption{Illustrations of our two primary evaluation frameworks. Document stream samples are shown as stacks of documents of sizes reflecting the corresponding sample sizes.}
  \label{fig:evaluation}
\end{figure*}

We conduct evaluations on the two primary tasks illustrated in~\autoref{fig:evaluation}. First, our ranking evaluation described in~\autoref{sec:ranking} is motivated by information retrieval needs. Although ranking is not the focus of this paper, it provides an assessment of the quality of the learned embedding in terms of similarity judgements and facilitates comparison with the baseline model~IUR. In addition, we use the ranking evaluation to monitor training using development data disjoint from the test data used for the final evaluation. Second, we introduce an account linking evaluation framework in~\autoref{sec:al} inspired by similar evaluations used for speaker recognition~\cite{doddington2000nist,van2007introduction}.
Both evaluations involve setting up two sets of samples as described below, the {\em queries} and the {\em targets}.
For each query, there is exactly one target drawn from the same document stream. Roughly speaking, both evaluations involve matching targets with their corresponding queries.

\subsection{MUD: a Web-scale training dataset}\label{sec:data}
Reddit is currently one of the most popular social media platforms, where anonymous users interact primarily by posting {\em comments} to discussion threads. Together with its text content, each comment is labeled by its publication time and the subreddit to which it was posted, a categorical feature roughly indicating its topic.

We construct a dataset consisting of 300~million Reddit posts from 1~million users published over an entire year to be used to train our proposed model. This {\em Million User Dataset} (MUD) consists of all posts by authors who published at least 100 and at most 1000~posts between July 2015 and June 2016, where the lower bound ensures a sufficiently long history from which to sample, and the upper bound is intended to reduce the impact of bot and spam accounts. We obtained the data by drawing from the existing Pushshift Reddit corpus~\cite{baumgartner2020pushshift}. Some further statistics of~MUD are shown in~\autoref{tab:datastats}.

\begin{table}
\begin{center}
\small
\begin{tabular}{cccccc}
        \toprule
        \textbf{Model} &\textbf{Features}&\textbf{Loss}&\textbf{MRR}  & \textbf{R@4} & \textbf{R@8} \\
        \midrule
        Prop & TPS & Top-$k$ &      \textbf{0.637} &  \textbf{0.709} &  \textbf{0.765}  \\
        Prop & TPS & Triplet &        0.634     &   0.702 &      0.762 \\
        \midrule
        Prop & TP & Top-$k$ & 0.450 &   \textbf{0.522} &         \textbf{0.595}  \\
        Prop & TP & Triplet &   \textbf{0.452} &   0.520 &         0.591  \\
        Prop & T & Triplet &   0.372 &   0.439 &         0.512  \\
        \midrule
        IUR  & TPS & Arcface & 0.520  & 0.590 & 0.650 \\
        IUR & T & Arcface & 0.200 & 0.240 & 0.290 \\
        \bottomrule
    \end{tabular}
    \caption{Ranking results for the proposed model (Prop) and the baseline (IUR), both trained and evaluated using various combinations of text content (T), publication time (P), and subreddit (S). IUR results are reported from
      \newcite{andrews-bishop-2019-learning}.}
\label{tab:finalresults}
\end{center}
\end{table}

\subsection{Ranking evaluation}\label{sec:ranking}
As shown in~\autoref{fig:ranking}, the ranking experiment consists of ranking the targets by similarity to each query.
For compatibility, we mimic the experimental setup from \newcite{andrews-bishop-2019-learning}, which proposes {\em separate} sets of queries and targets to be used for training and testing. We adopt the training split for validation and the testing split for evaluation, although we {\em train} our model on MUD (see~\autoref{sec:data}). We select hyperparameters based on dev split performance (see~\autoref{sec:hyper}). The test split consists of samples, each of size exactly~16, although we train the proposed model using samples from MUD of {\em varying} sizes as described in~\autoref{sec:sampling}. Note that the posts comprising MUD {\em precede} those of both IUR splits in publication time, ensuring that our training data is disjoint from IUR's test data.
Of the 111,396 authors contributing to the test split, 69,275 or 62\% contribute to the IUR training split. In contrast, MUD has only 39,529 users in common with the test split, a significantly smaller overlap than IUR. In principle, the increase in novel users at test time puts the proposed model at a disadvantage because it places more importance on generalization to novel users.
 
We report recall-at-$k$ (R@$k$) and mean reciprocal rank (MRR), calulated exactly as in~\citet{andrews-bishop-2019-learning}.
MRR is the expected value of the reciprocal of the position of the correct target in the ranked list. R@$k$ is the probability that the unique target composed by the same author as a given query appears in the top $k$~ranked results.
We limit ourselves to R@4 and R@8 as proxies for the ``first page'' of search results returned to a user issuing a query.

The results of this evaluation calculated with the test split are shown in~\autoref{tab:finalresults}.
Note that the full version of the proposed model significantly outperforms the previously published state-of-the-art.\footnote{A paired sign test of the differences in ranking between IUR and the proposed model is significant at the $p < 10^{-15}$ level.}
We conclude that although the angular margin loss used by~\citet{andrews-bishop-2019-learning} is considered state-of-the-art, the simpler triplet loss outperforms it, most likely because it admits the use of a considerably larger dataset.
We remark that the models trained with top-$k$ performed only slightly better than those trained with triplet loss, an
observation consistent with recent findings that when matching experimental conditions, the choice of ranking loss is less important than previously believed~\cite{musgrave2020metric}.

\begin{figure}[ht]
    \centering
    \begin{tikzpicture}[xscale=.65,yscale=.85]
\draw[xstep=1,ystep=1,gray,very thin] (-0.1,-0.1) grid (9.1,4.1);
\draw[thick,->] (-0.2,0) -- (9.2,0) node[below] {Hour};
\draw[thick,->] (0,-0.2) -- (0,4.2) node[left] {R@8};
\foreach \x in {5,10,15,20,25,30,35,40}
  \draw (\x/5.0,1pt) -- (\x/5.0,-1pt) node[anchor=north] {$\x$};
\foreach \y in {0.25,0.5,0.75}
  \draw (-1pt,\y*4) -- (1pt,\y*4) node[anchor=east] {$\y$};
\definecolor{localColor0}{RGB}{51,34,136}
\draw[thick,localColor0]  (0.0,2.11616)  -- (0.2,2.30176)  -- (0.4,2.41488)  -- (0.6,2.48976)  -- (0.8,2.54976)  -- (1.0,2.60336)  -- (1.2,2.63664)  -- (1.4,2.66736)  -- (1.6,2.69392)  -- (1.8,2.73936)  -- (2.0,2.7496)  -- (2.2,2.77952)  -- (2.4,2.8016)  -- (2.6,2.81136)  -- (2.8,2.82688)  -- (3.0,2.84512)  -- (3.2,2.85952)  -- (3.4,2.86896)  -- (3.6,2.88336)  -- (3.8,2.8944)  -- (4.0,2.90096)  -- (4.2,2.90912)  -- (4.4,2.91728)  -- (4.6,2.928)  -- (4.8,2.93808)  -- (5.0,2.9488)  -- (5.2,2.94576)  -- (5.4,2.95232)  -- (5.6,2.96128)  -- (5.8,2.96912)  -- (6.0,2.9696)  -- (6.2,2.98464)  -- (6.4,2.98304)  -- (6.6,2.99712)  -- (6.8,3.00816)  -- (7.0,2.99504)  -- (7.2,3.0048)  -- (7.4,3.00336)  -- (7.6,3.01488)  -- (7.8,3.0192)  -- (8.0,3.02448)  -- (8.2,3.02496)  -- (8.4,3.04464)  -- (8.6,3.04784)  -- (8.8,3.04912)  -- (9.0,3.04704) node[right] {TPS};
\definecolor{localColor3}{RGB}{17,119,51}
\draw[thick,localColor3]  (0.0,1.16944)  -- (0.2,1.42224)  -- (0.4,1.57728)  -- (0.6,1.67568)  -- (0.8,1.76544)  -- (1.0,1.82944)  -- (1.2,1.8936)  -- (1.4,1.94544)  -- (1.6,1.97136)  -- (1.8,2.02288)  -- (2.0,2.03968)  -- (2.2,2.07024)  -- (2.4,2.0864)  -- (2.6,2.12272)  -- (2.8,2.13312)  -- (3.0,2.14032)  -- (3.2,2.1624)  -- (3.4,2.18544)  -- (3.6,2.18864)  -- (3.8,2.2024)  -- (4.0,2.21536)  -- (4.2,2.2136)  -- (4.4,2.22768)  -- (4.6,2.24112)  -- (4.8,2.25088)  -- (5.0,2.25456)  -- (5.2,2.264)  -- (5.4,2.28144)  -- (5.6,2.28928)  -- (5.8,2.28896)  -- (6.0,2.29936)  -- (6.2,2.31152)  -- (6.4,2.34416)  -- (6.6,2.34992)  -- (6.8,2.34928)  -- (7.0,2.3488)  -- (7.2,2.35664)  -- (7.4,2.35888)  -- (7.6,2.35216)  -- (7.8,2.36112)  -- (8.0,2.36208)  -- (8.2,2.36528)  -- (8.4,2.35744)  -- (8.6,2.36032)  -- (8.8,2.36352)  -- (9.0,2.36768) node[below right] {TP};
\definecolor{localColor7}{RGB}{136,34,85}
\draw[thick,localColor7,dashed]  (0,2.6) -- (9.0,2.6) node[right] {IUR};
\end{tikzpicture}
    \caption{Test results after every hour of training for the proposed model using combinations of text (T), publication time (P), and subreddit (S) features as shown. The baseline result 0.65 shown as a dashed line is reported from~\newcite{andrews-bishop-2019-learning} and corresponds with a model availing of all three features.}
    \label{fig:learning}
\end{figure}
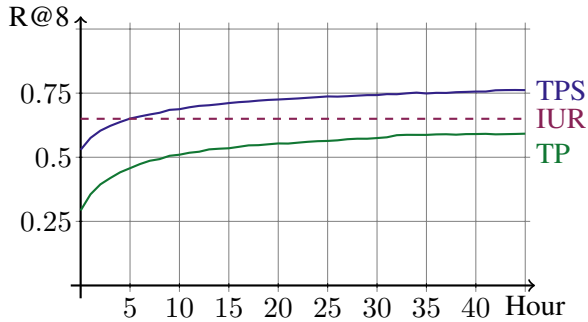

In addition, \autoref{fig:learning} shows the results of the evaluation performed after every hour of training.\footnote{Although we generated \autoref{fig:learning} {\em post-hoc} using test data, we did not use test data for model or hyperparameter selection.}
Note that after only six hours of training the full model outperforms the baseline.
\autoref{fig:learning} also shows the learning curve for an ablation of our model that eliminates the subreddit feature. We observe that this ablated model performs almost as well as the full-featured baseline, which suggests that the proposed approach may be effective in domains where only text and timestamps are available.

\subsection{A new framework for account linking}\label{sec:al}

While the ranking experiments in~\S\ref{sec:ranking} were designed to measure the quality of the learned embedding, they do not directly measure task performance: moderation applications require {\em decisions} rather than {\em rankings}. To this end, we propose an account linking benchmark modeled after the problem of enforcing account bans, in which a fixed number of accounts are linked against novel accounts at test-time. Compared to the ranking experiments, the key difference is that we introduce a {\em distinguished} subset of authors from which we have accumulated a significant number of previously published documents to serve as queries. The procedure is illustrated in~\autoref{fig:linking}.

Because the subreddit feature serves as a proxy for topic, restricting to a single subreddit results in a more challenging problem by increasing the likelihood that the comments considered deal with similar topics. To this end, we repeat the following procedure for each of the five most popular subreddits. Each result of the experiment reported in~\Cref{tab:linking,tab:known} is the average over the five subreddits of the corresponding results calculated using those subreddits individually.

Given a specified subreddit, we first randomly select 100~distinguished accounts, each publishing at least 100 posts to that subreddit in November~2016. The queries in the experiment consist of the 100~most recently published posts to the subreddit by each of the distinguished accounts in November~2016. In addition, the distinguished accounts must have published at least $16$~posts to the subreddit between December~2016 and May~2017 to serve as the corresponding targets, as described below.

Next we randomly select $4900$~accounts distinct from the distinguished accounts, each publishing at least $16$~posts to the subreddit between December 2016 and May 2017. The targets in the experiment consist of the 4~most recently published posts to the subreddit by each of the 5000~accounts.

\vspace{5pt} \noindent \textbf{Performance metrics}. For every query and target, each model considered returns a score, with smaller scores associated with a higher likelihood that the query and the target have the same author.
For example, the proposed model returns the distance between their embeddings under the model.
A decision rule to predict an author match is obtained by thresholding this score with respect to a chosen \emph{operating point}.
In production settings, one adjusts the operating point to obtain acceptable rates of false positives and false negatives.
In our running application of ban enforcement, these types of errors correspond respectively with mistakenly banning an innocent user and failing to ban a new account of a banned user.
Because the severity of these types errors are different, 
we consider the {\em detection cost function}
\[C_{\text{det}}= \pi C^{-} P^{-} +  \left(1 - \pi\right) C^{+} P^{+}\]
proposed by~\citet{van2007introduction}, where $P^{-}$ and $P^{+}$ are empirical probabilities of false negatives and false positives, $C^{-}$ and $C^{+}$ are the costs of false negatives and false positives, and $\pi$ is the {\em a priori} probability of a match.
We take $\pi=0.05$ and we set
$C^{-} = 1$ and $C^{+} = 2$, reflecting our presumption that banning an innocent account is more severe than failing to recognize a banned user. 
Our choices of $C^{-}$ and $C^{+}$ are only meant to reflect the asymmetric nature of the problem, although
in practice these costs would be highly platform-specific.

We report the minimum value of $C_\text{det}$ over all operating points (minDCF) and the value of $P^{+}$ at the operating point for which $P^{-}=P^{+}$, also known as the {\em equal error rate} (EER).

\begin{table}[ht!]
\centering
\begin{tabular}{cccc}
\toprule
\textbf{Model} & \textbf{Training}  & \textbf{EER} &  \textbf{minDCF} \\
&\textbf{Length}\\
\midrule
TF-IDF &     --       & 0.341 &   0.971 \\
Universal &    --     & 0.363 &   0.981 \\
IUR & 16              & 0.247 &   0.999 \\
\midrule
TP&1--8               & 0.169 &   0.848 \\
TP&1--16              & 0.132 &   0.792 \\
\bottomrule
\end{tabular}
\caption{Linking evaluation, averaged over 5 subreddits. The proposed model was trained using triplet loss. Smaller scores are better for both metrics.}
\label{tab:linking}
\end{table}

\vspace{5pt} \noindent \textbf{Baseline models}. We compare the proposed method with three baselines.
First we consider TF-IDF vector representations of the concatenated text content of a sample, which are compared using cosine similarity.
Next, we consider universal sentence encodings~\cite{cer2018universal}, which are compared using angular distance. We experimented with two versions of this baseline, namely embedding the concatenation of the text content of the documents in a sample, and averaging the embeddings of the individual documents. Since we found the concatenated version to perform better, we only report on this variation. Finally, we consider
IUR~\cite{andrews-bishop-2019-learning}. Because this model only embeds samples of size~16, we pad samples containing fewer than 16~posts. To handle samples containing more than 16~posts,  we organize the sample into contiguous groups of at most 16~posts, apply the embedding to each group, and average the embeddings.

\vspace{5pt} \noindent \textbf{Results}. \autoref{tab:linking} compares the linking performance of the three baseline models along with two of variations of the proposed model arising from varying the sizes of the training samples. Note that both variations of the proposed model outperform the baselines. A further variation on this experiment is reported in~\autoref{tab:known} in which the queries are drawn from the training dataset, better reflecting the context of the motivating example.

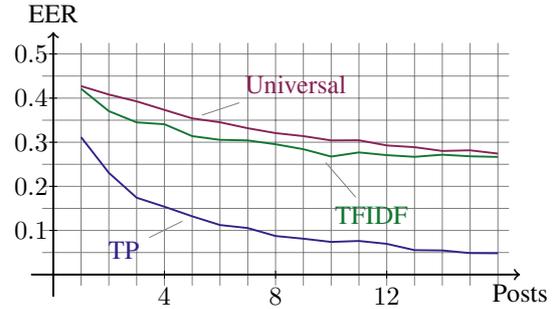
\begin{figure}[ht!]
    \centering
    \scalebox{.9}{\begin{tikzpicture}[scale=1.625]
\draw[xstep=0.25,ystep=.2,gray,very thin] (-0.1,-0.1) grid (4.1,2.1);
\draw[thick,->] (-0.2,0) -- (4.2,0) node[below] {Posts};
\draw[thick,->] (0,-0.2) -- (0,2.2) node[above] {EER};
\foreach \x in {4,8,12}
  \draw (\x/4.0,1pt) -- (\x/4.0,-1pt) node[below] {$\x$};
\foreach \y/\ytext in {0.1,0.2,0.3,0.4,0.5}
  \draw (-1pt,4.0*\y) -- (1pt,4.0*\y) node[left] {$\y$};
\definecolor{localColor0}{RGB}{51,34,136}
\definecolor{localColor3}{RGB}{17,119,51}
\definecolor{localColor7}{RGB}{136,34,85}

\draw[thick,localColor7]  (0.25,1.7089) -- (0.5,1.6324) -- (0.75,1.5705) -- (1.0,1.4931) -- (1.25,1.4163) node[pin=20:Universal] {}-- (1.5,1.3809) -- (1.75,1.3269) -- (2.0,1.2828) -- (2.25,1.2555) -- (2.5,1.2165) -- (2.75,1.2178) -- (3.0,1.1709) -- (3.25,1.1552) -- (3.5,1.1209) -- (3.75,1.1267) -- (4.0,1.0970);

\draw[thick,localColor0]  (0.25,1.2461) -- (0.5,0.9225) -- (0.75,0.6969) -- (1.0,0.6152) -- (1.25,0.5283) node[pin=200:TP] {} -- (1.5,0.4493) -- (1.75,0.4218) -- (2.0,0.3495) -- (2.25,0.3252) -- (2.5,0.2959) -- (2.75,0.3048) -- (3.0,0.2787) -- (3.25,0.2215) -- (3.5,0.2189) -- (3.75,0.1948) -- (4.0,0.1934);

\draw[thick,localColor3]  (0.25,1.6851) -- (0.5,1.4820) -- (0.75,1.3800) -- (1.0,1.3632) -- (1.25,1.2554) -- (1.5,1.2217) -- (1.75,1.2165) -- (2.0,1.1815) -- (2.25,1.1367) -- node[pin=280:TFIDF] {} (2.5,1.0701) -- (2.75,1.1077) -- (3.0,1.0830) -- (3.25,1.0677) -- (3.5,1.0868) -- (3.75,1.0725) -- (4.0,1.0668);

\end{tikzpicture}}
    \caption{Equal error rate (EER) as the length of the target samples varies. Results are averaged over 5~subreddits. Smaller is better for EER. IUR is omitted as it does not admit variable-sized samples.}
    \label{fig:eer}
\end{figure}

\begin{figure}[ht!]
    \centering
    \input{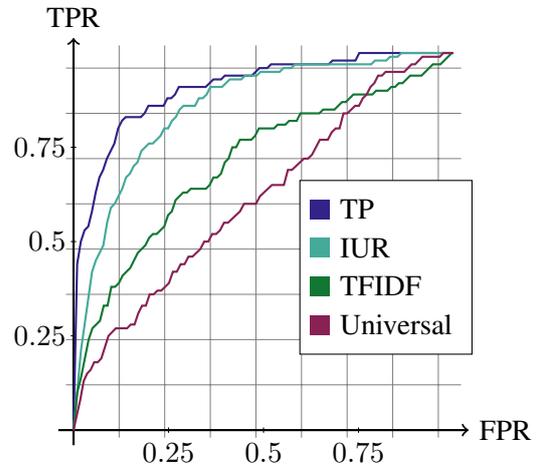}
    \caption{Reciever operator curves of models evaluated on a single subreddit and averaged over the five most popular subreddits. A point on a ROC near the northwest corner corresponds with an operating point with low FPR and high TPR, so a model is better if its ROC stretches further to the northwest.}
    \label{fig:roc}
\end{figure}

\autoref{fig:eer} shows the effect of the sizes of the targets on linking performance. Note that performance rapidly improves with larger samples, relative to the baselines. 
This trend is promising for our motivating application of ban enforcement, where it is desirable to recognize banned users as early as possible.
\autoref{fig:roc} shows receiver operator curves (ROC), which plot false positive rates against true positive rates as the operating points vary.

\subsection{Embeddings of variable-sized samples}\label{sec:varlen}

Our experiments in~\autoref{sec:al} show that linking samples from newly created accounts to those of distinguished authors is more successful when using as much historical data from the distinguished authors as possible. However, computational constraints typically inhibit embedding full account histories during training. Instead, in~\autoref{sec:al} we embed large samples of distinguished authors' histories using models trained on samples of sizes up to a maximum tractable length~$S$. We take $S=16$ in our experiments as described in~\autoref{sec:sampling}. Here, we examine the ability of a model trained on samples of sizes at most~$S$ to generalize to samples of sizes greater than~$S$.

We also compare to a further baseline that averages single-post embeddings produced by a variation of the proposed model trained on single posts, which we denote by Avg. This is in contrast with the proposed model, which aggregates embeddings of multiple posts using an attention mechanism.

\autoref{tab:variableresults} shows the ranking performance of a number of variations of the proposed model trained with triplet loss and all features (TPS) on samples of fixed or varying sizes as specified.
These results demonstrate that a model trained on variable-sized samples appears to generalize well to much longer samples. We observe substantially better performance compared to the simple averaging baseline, and only a slight decrease in performance compared to the fixed length models as the evaluation sample size increases beyond the lengths seen at training time.

\begin{table}[ht!]
\begin{center}
\begin{tabular}{ccccc}
    \toprule
    \textbf{Train} &\textbf{Test} &\textbf{Model} &\textbf{MRR} & \textbf{R@4}\\
    \textbf{Size} &\textbf{Size}\\
    \midrule
  1  & 8 &Avg & 0.205 & 0.243 \\
  1--8   & 8&Prop & 0.349 & 0.411 \\
 8  & 8   &Prop& 0.358 & 0.420 \\
 \midrule
 1   & 12  &Avg & 0.282 & 0.330 \\
  1--8   & 12 &Prop& 0.497 & 0.568 \\
 12 & 12&Prop  & 0.516  & 0.589  \\
 \midrule
  1  & 16  &Avg & 0.347 & 0.397  \\
 1--8   & 16&Prop & 0.603 & 0.672  \\
 16 & 16&Prop  & 0.632 & 0.700 \\
    \bottomrule
\end{tabular} \\
\caption{Performance of the proposed model (Prop) and a simple averaging baseline (Avg), both trained and evaluated on fixed or variable-sized samples as shown.}
\label{tab:variableresults}
\end{center}
\end{table}
\subsection{Selecting the distribution of sizes}\label{sec:beta_expt}

As mentioned in~\autoref{sec:sampling}, we use $\mathrm{Beta}\left(3,1\right)$ to select sample sizes during training. We hypothesize that a negatively skewed distribution tends to improve training efficiency by supplying longer samples most of the time, while retaining the ability to handle shorter samples.
To evaluate this claim, we investigate several distributions of varying degrees of negative skew.

\autoref{tab:betaresults} shows the ranking performance of variations of the proposed model
trained on samples of sizes varying between 1 and 16 posts and evaluated on samples of size~16. These models differ only in the distribution used to select sample sizes. Indeed, the negatively skewed distributions {\em do} improve ranking performance over the uniform distribution, although the choice of negatively skewed distribution appears to be mostly immaterial.

\begin{table}[ht]
\begin{center}
\begin{tabular}{l c  c  c c}
    \toprule
    \textbf{Model}& \textbf{Skew} &\textbf{MRR} & \textbf{R@4}\\
    \midrule
 $\mathrm{Unif}\left(0,1\right)$  & 0  & 0.604 & 0.673   \\
 $\mathrm{Beta}\left(2,1\right)$ & -0.566 & 0.626  & 0.692       \\
$\mathrm{tPois}\left(16\right)$   & -0.786 & 0.632    & 0.703  \\
$\mathrm{Beta}\left(3,1\right)$   & -0.861 & 0.633   & 0.701  \\
$\mathrm{Beta}\left(4,1\right)$   & -1.049 & 0.633    & 0.704  \\
    \bottomrule
\end{tabular} \\
\end{center}
\caption{Ranking evaluations of models trained on samples of sizes selected with various distributions. $\mathrm{tPois}\left(16\right)$
is derived from the Poisson distribution with mean~16 by truncating its support to $\left\{1,2,\ldots,16\right\}$.}
\label{tab:betaresults}
\end{table} 

\section{Future work}\label{sec:conclusion}

This work motivates a number of interesting research questions. 
First, the proposed model makes use of publication times, but only avails of the hour of the day.
It would be interesting to examine continuous-time variants of our encoder that incorporate relative time differences between actions when aggregating their embeddings, in light of the fact that patterns of user activity might be highly discriminative. For example, bots and spammers typically post at certain times of day and with particular frequencies.
Separately, the proposed data augmentation methods we use to  handle variable-sized
samples may also be applicable in other settings, such as multi-document summarization~\cite{liu2019hierarchical}.
Finally, the scores we use to determine author matches could be \emph{calibrated}, providing confidence estimates associated with the account linking decisions.

To our knowledge, this work is the first to demonstrate the {\em feasibility} of a general-purpose account linking framework at web scale. Indeed, \autoref{fig:eer} shows that performance improves as the size of the target increases, suggesting a speed-accuracy trade-off that can be tuned for different application settings. Expanding on an idea above, if confidence estimates were available, they could be used to inform the necessary sample sizes to achieve an acceptable level of risk.

Finally, we note that the generality of the proposed approach make it potentially applicable to a wide range of applications, including source code attribution~\cite{burrows2009application,yang2017authorship,kalgutkar2019code}, plagiarism detection~\cite{potthast2010evaluation,meuschke2018hyplag,foltynek2019academic}, and authorship attribution in collaborative documents~\cite{flock2014wikiwho,dauber2017stylometric}.

\bibliographystyle{acl_natbib}
\bibliography{paper}

\begin{thebibliography}{37}
\expandafter\ifx\csname natexlab\endcsname\relax\def\natexlab#1{#1}\fi

\bibitem[{{Alvari} et~al.(2019){Alvari}, {Sarkar}, and
  {Shakarian}}]{violentExtermeism}
H.~{Alvari}, S.~{Sarkar}, and P.~{Shakarian}. 2019.
\newblock Detection of violent extremists in social media.
\newblock In \emph{2019 2nd International Conference on Data Intelligence and
  Security (ICDIS)}, pages 43--47.

\bibitem[{Andrews and Bishop(2019)}]{andrews-bishop-2019-learning}
Nicholas Andrews and Marcus Bishop. 2019.
\newblock \href {https://doi.org/10.18653/v1/D19-1178} {Learning invariant
  representations of social media users}.
\newblock In \emph{Proceedings of the 2019 Conference on Empirical Methods in
  Natural Language Processing and the 9th International Joint Conference on
  Natural Language Processing (EMNLP-IJCNLP)}, pages 1684--1695, Hong Kong,
  China. Association for Computational Linguistics.

\bibitem[{Baumgartner et~al.(2020)Baumgartner, Zannettou, Keegan, Squire, and
  Blackburn}]{baumgartner2020pushshift}
Jason Baumgartner, Savvas Zannettou, Brian Keegan, Megan Squire, and Jeremy
  Blackburn. 2020.
\newblock \href {http://arxiv.org/abs/2001.08435} {The pushshift reddit
  dataset}.

\bibitem[{Broniatowski et~al.(2018)Broniatowski, Jamison, Qi, AlKulaib, Chen,
  Benton, Quinn, and Dredze}]{broniatowski2018weaponized}
David~A Broniatowski, Amelia~M Jamison, SiHua Qi, Lulwah AlKulaib, Tao Chen,
  Adrian Benton, Sandra~C Quinn, and Mark Dredze. 2018.
\newblock Weaponized health communication: Twitter bots and russian trolls
  amplify the vaccine debate.
\newblock \emph{American journal of public health}, 108(10):1378--1384.

\bibitem[{Burrows et~al.(2009)Burrows, Uitdenbogerd, and
  Turpin}]{burrows2009application}
Steven Burrows, Alexandra~L Uitdenbogerd, and Andrew Turpin. 2009.
\newblock Application of information retrieval techniques for source code
  authorship attribution.
\newblock In \emph{International Conference on Database Systems for Advanced
  Applications}, pages 699--713. Springer.

\bibitem[{Caliskan et~al.(2017)Caliskan, Bryson, and
  Narayanan}]{caliskan2017semantics}
Aylin Caliskan, Joanna~J Bryson, and Arvind Narayanan. 2017.
\newblock Semantics derived automatically from language corpora contain
  human-like biases.
\newblock \emph{Science}, 356(6334):183--186.

\bibitem[{Cer et~al.(2018)Cer, Yang, Kong, Hua, Limtiaco, St.~John, Constant,
  Guajardo-Cespedes, Yuan, Tar, Strope, and Kurzweil}]{cer2018universal}
Daniel Cer, Yinfei Yang, Sheng-yi Kong, Nan Hua, Nicole Limtiaco, Rhomni
  St.~John, Noah Constant, Mario Guajardo-Cespedes, Steve Yuan, Chris Tar,
  Brian Strope, and Ray Kurzweil. 2018.
\newblock \href {https://doi.org/10.18653/v1/D18-2029} {Universal sentence
  encoder for {E}nglish}.
\newblock In \emph{Proceedings of the 2018 Conference on Empirical Methods in
  Natural Language Processing: System Demonstrations}, pages 169--174,
  Brussels, Belgium. Association for Computational Linguistics.

\bibitem[{Corbett-Davies and Goel(2018)}]{corbett2018measure}
Sam Corbett-Davies and Sharad Goel. 2018.
\newblock The measure and mismeasure of fairness: A critical review of fair
  machine learning.
\newblock \emph{arXiv preprint arXiv:1808.00023}.

\bibitem[{Daelemans et~al.(2019)Daelemans, Kestemont, Manjavacas, Potthast,
  Rangel, Rosso, Specht, Stamatatos, Stein, Tschuggnall, Wiegmann, and
  Zangerle}]{pan:2019b}
Walter Daelemans, Mike Kestemont, Enrique Manjavacas, Martin Potthast,
  Francisco Rangel, Paolo Rosso, G{\"{u}}nther Specht, Efstathios Stamatatos,
  Benno Stein, Michael Tschuggnall, Matti Wiegmann, and Eva Zangerle. 2019.
\newblock \href {http://ceur-ws.org/Vol-2380/} {{Overview of PAN 2019: Author
  Profiling, Celebrity Profiling, Cross-domain Authorship Attribution and Style
  Change Detection}}.
\newblock In \emph{10th International Conference of the CLEF Association (CLEF
  2019)}. Springer.

\bibitem[{Dauber et~al.(2017)Dauber, Overdorf, and
  Greenstadt}]{dauber2017stylometric}
Edwin Dauber, Rebekah Overdorf, and Rachel Greenstadt. 2017.
\newblock Stylometric authorship attribution of collaborative documents.
\newblock In \emph{International Conference on Cyber Security Cryptography and
  Machine Learning}, pages 115--135. Springer.

\bibitem[{Davidson et~al.(2017)Davidson, Warmsley, Macy, and
  Weber}]{davidson2017automated}
Thomas Davidson, Dana Warmsley, Michael Macy, and Ingmar Weber. 2017.
\newblock Automated hate speech detection and the problem of offensive
  language.
\newblock In \emph{Eleventh International AAAI Conference on Web and Social
  Media}.

\bibitem[{Devlin et~al.(2019)Devlin, Chang, Lee, and
  Toutanova}]{devlin2018bert}
Jacob Devlin, Ming-Wei Chang, Kenton Lee, and Kristina Toutanova. 2019.
\newblock \href {https://doi.org/10.18653/v1/N19-1423} {{BERT}: Pre-training of
  deep bidirectional transformers for language understanding}.
\newblock In \emph{Proceedings of the 2019 Conference of the North {A}merican
  Chapter of the Association for Computational Linguistics: Human Language
  Technologies, Volume 1 (Long and Short Papers)}, pages 4171--4186,
  Minneapolis, Minnesota. Association for Computational Linguistics.

\bibitem[{Doddington et~al.(2000)Doddington, Przybocki, Martin, and
  Reynolds}]{doddington2000nist}
George~R Doddington, Mark~A Przybocki, Alvin~F Martin, and Douglas~A Reynolds.
  2000.
\newblock The nist speaker recognition evaluation--overview, methodology,
  systems, results, perspective.
\newblock \emph{Speech communication}, 31(2-3):225--254.

\bibitem[{Fan(2012)}]{fan2012graph}
Wenfei Fan. 2012.
\newblock Graph pattern matching revised for social network analysis.
\newblock In \emph{Proceedings of the 15th International Conference on Database
  Theory}, pages 8--21.

\bibitem[{Fl{\"o}ck and Acosta(2014)}]{flock2014wikiwho}
Fabian Fl{\"o}ck and Maribel Acosta. 2014.
\newblock Wikiwho: Precise and efficient attribution of authorship of
  revisioned content.
\newblock In \emph{Proceedings of the 23rd international conference on World
  wide web}, pages 843--854.

\bibitem[{Folt{\`y}nek et~al.(2019)Folt{\`y}nek, Meuschke, and
  Gipp}]{foltynek2019academic}
Tom{\'a}{\v{s}} Folt{\`y}nek, Norman Meuschke, and Bela Gipp. 2019.
\newblock Academic plagiarism detection: a systematic literature review.
\newblock \emph{ACM Computing Surveys (CSUR)}, 52(6):1--42.

\bibitem[{Ishihara(2011)}]{ishihara2011forensic}
Shunichi Ishihara. 2011.
\newblock A forensic authorship classification in sms messages: A likelihood
  ratio based approach using n-gram.
\newblock In \emph{Proceedings of the Australasian Language Technology
  Association Workshop 2011}, pages 47--56.

\bibitem[{Kalgutkar et~al.(2019)Kalgutkar, Kaur, Gonzalez, Stakhanova, and
  Matyukhina}]{kalgutkar2019code}
Vaibhavi Kalgutkar, Ratinder Kaur, Hugo Gonzalez, Natalia Stakhanova, and Alina
  Matyukhina. 2019.
\newblock Code authorship attribution: Methods and challenges.
\newblock \emph{ACM Computing Surveys (CSUR)}, 52(1):1--36.

\bibitem[{Kim et~al.(2020)Kim, Kim, Cho, and Kwak}]{kim2020proxy}
Sungyeon Kim, Dongwon Kim, Minsu Cho, and Suha Kwak. 2020.
\newblock \href {http://arxiv.org/abs/2003.13911} {Proxy anchor loss for deep
  metric learning}.

\bibitem[{Kudo(2018)}]{kudo2018subword}
Taku Kudo. 2018.
\newblock \href {https://doi.org/10.18653/v1/P18-1007} {Subword regularization:
  Improving neural network translation models with multiple subword
  candidates}.
\newblock In \emph{Proceedings of the 56th Annual Meeting of the Association
  for Computational Linguistics (Volume 1: Long Papers)}, pages 66--75,
  Melbourne, Australia. Association for Computational Linguistics.

\bibitem[{Lan et~al.(2019)Lan, Chen, Goodman, Gimpel, Sharma, and
  Soricut}]{lan2019albert}
Zhenzhong Lan, Mingda Chen, Sebastian Goodman, Kevin Gimpel, Piyush Sharma, and
  Radu Soricut. 2019.
\newblock Albert: A lite bert for self-supervised learning of language
  representations.
\newblock In \emph{International Conference on Learning Representations}.

\bibitem[{Liu and Lapata(2019)}]{liu2019hierarchical}
Yang Liu and Mirella Lapata. 2019.
\newblock \href {https://doi.org/10.18653/v1/P19-1500} {Hierarchical
  transformers for multi-document summarization}.
\newblock In \emph{Proceedings of the 57th Annual Meeting of the Association
  for Computational Linguistics}, pages 5070--5081, Florence, Italy.
  Association for Computational Linguistics.

\bibitem[{Lu et~al.(2019)Lu, Xu, Zhang, Duan, and Mei}]{Lu_2019_ICCV}
Jing Lu, Chaofan Xu, Wei Zhang, Ling-Yu Duan, and Tao Mei. 2019.
\newblock Sampling wisely: Deep image embedding by top-k precision
  optimization.
\newblock In \emph{The IEEE International Conference on Computer Vision
  (ICCV)}.

\bibitem[{McCoy et~al.(2019)McCoy, Pavlick, and Linzen}]{mccoy2019right}
Tom McCoy, Ellie Pavlick, and Tal Linzen. 2019.
\newblock \href {https://doi.org/10.18653/v1/P19-1334} {Right for the wrong
  reasons: Diagnosing syntactic heuristics in natural language inference}.
\newblock In \emph{Proceedings of the 57th Annual Meeting of the Association
  for Computational Linguistics}, pages 3428--3448, Florence, Italy.
  Association for Computational Linguistics.

\bibitem[{McInnes et~al.(2020)McInnes, Healy, and Melville}]{mcinnes2020umap}
Leland McInnes, John Healy, and James Melville. 2020.
\newblock \href {http://arxiv.org/abs/1802.03426} {Umap: Uniform manifold
  approximation and projection for dimension reduction}.

\bibitem[{Meuschke et~al.(2018)Meuschke, Stange, Schubotz, and
  Gipp}]{meuschke2018hyplag}
Norman Meuschke, Vincent Stange, Moritz Schubotz, and Bela Gipp. 2018.
\newblock Hyplag: a hybrid approach to academic plagiarism detection.
\newblock In \emph{The 41st International ACM SIGIR Conference on Research \&
  Development in Information Retrieval}, pages 1321--1324.

\bibitem[{Micikevicius et~al.(2017)Micikevicius, Narang, Alben, Diamos, Elsen,
  Garcia, Ginsburg, Houston, Kuchaiev, Venkatesh, and
  Wu}]{micikevicius2017mixed}
Paulius Micikevicius, Sharan Narang, Jonah Alben, Gregory Diamos, Erich Elsen,
  David Garcia, Boris Ginsburg, Michael Houston, Oleksii Kuchaiev, Ganesh
  Venkatesh, and Hao Wu. 2017.
\newblock \href {http://arxiv.org/abs/1710.03740} {Mixed precision training}.

\bibitem[{Musgrave et~al.(2020)Musgrave, Belongie, and
  Lim}]{musgrave2020metric}
Kevin Musgrave, Serge Belongie, and Ser-Nam Lim. 2020.
\newblock A metric learning reality check.
\newblock \emph{arXiv preprint arXiv:2003.08505}.

\bibitem[{Pennycook et~al.(2020)Pennycook, McPhetres, Zhang, Lu, and
  Rand}]{pennycook2020fighting}
Gordon Pennycook, Jonathon McPhetres, Yunhao Zhang, Jackson~G Lu, and David~G
  Rand. 2020.
\newblock Fighting covid-19 misinformation on social media: Experimental
  evidence for a scalable accuracy-nudge intervention.
\newblock \emph{Psychological science}, 31(7):770--780.

\bibitem[{Potthast et~al.(2010)Potthast, Stein, Barr{\'o}n-Cede{\~n}o, and
  Rosso}]{potthast2010evaluation}
Martin Potthast, Benno Stein, Alberto Barr{\'o}n-Cede{\~n}o, and Paolo Rosso.
  2010.
\newblock An evaluation framework for plagiarism detection.
\newblock In \emph{Coling 2010: Posters}, pages 997--1005.

\bibitem[{Schroff et~al.(2015)Schroff, Kalenichenko, and
  Philbin}]{schroff2015facenet}
Florian Schroff, Dmitry Kalenichenko, and James Philbin. 2015.
\newblock Facenet: A unified embedding for face recognition and clustering.
\newblock In \emph{Proceedings of the IEEE conference on computer vision and
  pattern recognition}, pages 815--823.

\bibitem[{Schwartz et~al.(2013)Schwartz, Tsur, Rappoport, and
  Koppel}]{schwartz2013authorship}
Roy Schwartz, Oren Tsur, Ari Rappoport, and Moshe Koppel. 2013.
\newblock Authorship attribution of micro-messages.
\newblock In \emph{Proceedings of the 2013 Conference on Empirical Methods in
  Natural Language Processing}, pages 1880--1891.

\bibitem[{Silvestri et~al.(2015)Silvestri, Yang, Bozzon, and
  Tagarelli}]{silvestri2015linking}
Giuseppe Silvestri, Jie Yang, Alessandro Bozzon, and Andrea Tagarelli. 2015.
\newblock Linking accounts across social networks: the case of stackoverflow,
  github and twitter.
\newblock In \emph{KDWeb}, pages 41--52.

\bibitem[{Van Der~Walt and Eloff(2018)}]{van2018using}
Est{\'e}e Van Der~Walt and Jan Eloff. 2018.
\newblock Using machine learning to detect fake identities: Bots vs humans.
\newblock \emph{IEEE Access}, 6:6540--6549.

\bibitem[{Van~Leeuwen and Br{\"u}mmer(2007)}]{van2007introduction}
David~A Van~Leeuwen and Niko Br{\"u}mmer. 2007.
\newblock An introduction to application-independent evaluation of speaker
  recognition systems.
\newblock In \emph{Speaker classification I}, pages 330--353. Springer.

\bibitem[{Wang et~al.(2019)Wang, Zhang, Huang, and Scott}]{wang2019crossbatch}
Xun Wang, Haozhi Zhang, Weilin Huang, and Matthew~R. Scott. 2019.
\newblock \href {http://arxiv.org/abs/1912.06798} {Cross-batch memory for
  embedding learning}.

\bibitem[{Yang et~al.(2017)Yang, Xu, Li, Guo, and Zhang}]{yang2017authorship}
Xinyu Yang, Guoai Xu, Qi~Li, Yanhui Guo, and Miao Zhang. 2017.
\newblock Authorship attribution of source code by using back propagation
  neural network based on particle swarm optimization.
\newblock \emph{PloS one}, 12(11):e0187204.

\end{thebibliography}
\clearpage
\appendix
\section{Hyperparameter selection}\label{sec:hyper}

Top-$k$ loss involves a number of hyperparameters. In addition to $k$ and the margin penalty~$m$, one must also select the number $n_{+}$ of targets from each class presented in every batch.
We found the values of $k,m,n_{+}$ suggested by~\cite{Lu_2019_ICCV} to not perform well in our setting. We therefore conducted a small grid search for these values, selecting the optimal configuration based on validation scores. We considered $k\in\left\{4, 8\right\}$, $n_+\in\left\{4, 16\right\}$, and $m\in\left\{0.05, 0.40\right\}$, resulting in $k = 4$, $n_+ = 8$, and $m = 0.25$.

In experiments with the triplet loss, we use a fixed margin $m = 0.2$. We arrived at this value through the grid search illustrated in~\autoref{tab:marginsearch}. We do not use dropout regularization, as previous work has shown dropout can degrade performance when training with larger datasets~\cite{lan2019albert}.

\begin{table}[ht]
    \centering
\begin{tabular}{ccc}
\toprule
Margin &   MRR &   R@8  \\
\midrule
0.1        & 0.59 & 0.73 \\
0.2       & \textbf{0.62} & \textbf{0.75} \\
0.3        & 0.59 & 0.74 \\
0.4        & 0.61 & 0.74 \\
0.5        & 0.60 & 0.73 \\
0.6        & 0.59 & 0.73 \\
0.8        & 0.58 & 0.72 \\
\bottomrule
\end{tabular}
    \caption{Validation ranking results for various triplet loss margins.}
    \label{tab:marginsearch}
\end{table}

\section{Implementation details}
\noindent{\bf Mixed precision.}
We use mixed precision training~\cite{micikevicius2017mixed}. This
reduces the GPU memory consumed at training time by about half through the use of half
precision floats, enables faster forward and backward pass computations, and
allows for a larger batch size.

\vspace{5pt}\noindent{\bf Multi-GPU training.} Using 2 V100 GPUs to train the model significantly speeds up the process by quadrupling the effective batch size. Our models have an average training time of around 72~hours.

\vspace{5pt}\noindent{\bf Simplified text encoding.}
We limit our convolutional text encoders to windows of 2, 3, and 4 subwords, excluding the largest window of 5 used by other models.
Surprisingly, this did not impact ranking performance, which suggests that a small receptive field is sufficient for purposes of comparing authorship. As further support for this claim, we also experimented with larger receptive fields than 5, which \emph{reduced} ranking performance.

\vspace{5pt}\noindent{\bf{Model Hyper-parameters.}}
Our reported models are trained with an embedding dimension of $D=1024$, 512~convolutional filters, and an attention mechanism producing outputs of dimension~512. Additionally the subword and subreddit embedding dimensions are both $N=512$.

\vspace{5pt}\noindent{\bf{Model Parameters}}
\autoref{tab:parameters} shows the numbers of parameters of the proposed model when using various combinations of text content (T), publication time (P), and subreddit (S).
\begin{table}[ht]
    \centering
\begin{tabular}{cc}
\toprule
Model  & Parameters \\
\midrule
T   & 44.0M \\
TP  & 44.1M \\
TPS & 45.5M \\
\bottomrule
\end{tabular}
    \caption{Numbers of parameters in various trained models.}
    \label{tab:parameters}
\end{table}

\section{Text encoding}
\label{sec:encode}
We consider two methods to encode raw text content into integer arrays, namely taking the integer values of the corresponding UTF\nobreakdash-8 encoded bytes directly, and using the SentencePiece unigram subword model~\cite{kudo2018subword}. SentencePiece tokenizes select character groupings according to a pretrained vocabulary of a specified size, which can be orders of magnitudes larger than the vocabulary of size~$2^8$ used by the byte encoding. For experiments conducted in \autoref{sec:experiments}, we used  SentencePiece with a vocabulary size of~$2^{16}$.

The potentially large disparity in vocabulary size between encoding methods can result in text encoded as integer arrays of significantly different lengths. In light of the need to truncate these arrays at training, we hypothesize that subword encoding with a large vocabulary results in better model performance as more textual information is captured after truncation.

 To evaluate this claim, we fit additional SentencePiece subword models with vocabulary sizes $2^{12}$ and $2^{14}$ on text content from MUD. \autoref{tab:subwordresults} provides ranking performance of variations of the proposed model trained with only the text feature on samples of fixed sized~16 and evaluated on samples of fixed sized~16. We remark that increases in vocabulary size {\em do} correspond with increases in ranking performance.
 We found that increasing the vocabulary size beyond~$2^{16}$ did {\em not} increase performance further.
 
\begin{table}
\begin{center}
\begin{tabular}{ccc  c  c c}
    \toprule
    \textbf{Model}&\textbf{Size}&\textbf{MRR} & \textbf{R@4} &\textbf{R@8} \\
    \midrule
  SP&$2^{16}$    & {0.372} & {0.439} & {0.512}   \\
 SP&$2^{14}$     & 0.352 & 0.412 & 0.483      \\
 SP&$2^{12}$   & 0.307 & 0.360   & 0.433  \\
 Byte&$2^{8} $   & 0.105  & 0.124  & 0.163       \\
    \bottomrule
\end{tabular} \\
\caption{Ranking performance of models using various subword vocabulary sizes. All models use SentencePiece (SP) or the byte encoding (Byte) described in~\autoref{sec:encode}.}
\label{tab:subwordresults}
\end{center}
\end{table}

\section{Further statistics about MUD}\label{sec:mud}
\autoref{tab:datastats} shows some further details of the MUD dataset introduced in~\autoref{sec:data}.

\begin{table}[ht]
\centering
\begin{tabular}{lr}
\toprule
Number of users contributing & 1,071,477\\\midrule
Number of posts & 321,659,421\\\midrule
Mean post length & 42.5~tokens\\\midrule
Mean number of posts\\contributed by a user & 300.2\\\midrule
Mean number of subreddits\\accessed by a user & 22.1\\\midrule
Mean number of months\\a user was active &9.9\\\midrule
Percentage of posts containing\\more than 64~tokens&17.37\%\\
\bottomrule
\end{tabular}
\caption{Some statistics of the Million User Dataset.}
\label{tab:datastats}
\end{table}

\section{Further experiments}

\begin{table}
    \centering
\begin{tabular}{cccc}
\toprule
\textbf{Model} & \textbf{Training}  & \textbf{EER} &  \textbf{minDCF} \\
&\textbf{Length}\\
\midrule 
TF-IDF                    & -- & 0.362 &   0.967 \\
Universal         & -- & 0.371 &   0.989 \\
IUR  &  16 & 0.223 &   0.951 \\
\midrule
TP  &  1--8 & 0.145 &   0.791 \\
TP  &  1--16 & 0.136 &   0.765 \\
\bottomrule
\end{tabular}
    \caption{Same experiment as~\autoref{tab:linking}, but with queries {\em observed} at training time.}
    \label{tab:known}
\end{table}

We repeat the linking experiment from~\autoref{sec:al} using queries that are observed at training time.
We expect this to improve performance, something which is confirmed in~\autoref{tab:known}.

\end{document}